# Mn$_{12}$-acetate film pattern generated by photolithography methods

K. Kim, D.M. Seo, J. Means, V. Meenakshi, W. Teizer[a]

Department of Physics, Texas A&M University, College Station, TX 77843-4242

H. Zhao, and K. R. Dunbar

Department of Chemistry, Texas A&M University, College Station, TX 77842-3012

*We demonstrate a straightforward way to lithographically fabricate Mn$_{12}$-acetate (Mn$_{12}$O$_{12}$(CH$_3$COO)$_{16}$(H$_2$O)$_4$·2CH$_3$COOH·4H$_2$O) thin film patterns on Si/SiO$_2$ surfaces, a significant step in light of the chemical volatility of this organic complex. Atomic force microscopy (AFM) images of the film patterns allow the determination of the pattern dimensions. X-ray photoelectron spectroscopy (XPS) data indicate that the patterned material is the intact Mn$_{12}$-acetate complex. Magnetic measurements of the Mn$_{12}$-acetate film confirm that the film properties are reminiscent of crystalline Mn$_{12}$-acetate, suggesting that this approach can be used to fabricate lithographically patterned devices of Mn$_{12}$-acetate.*

[a]Electronic address: teizer@tamu.edu

There have been few efforts to fabricate ordered patterns of single molecule magnets (SMM),[1-3] the most thoroughly studied of which is $Mn_{12}$-acetate ($Mn_{12}O_{12}(CH_3COO)_{16}(H_2O)_4·2CH_3COOH·4H_2O$).[4-9] The unusual magnetic properties of these molecules, in particular the stepwise quantum magnetic hysteresis loops observed for the individual molecules,[10-14] have led to speculation that such clusters are potential candidates for high-density magnetic storage and quantum computing applications. The fabrication of artificially organized patterns of these molecules is an important issue in this area because isolated single magnetic domains or aggregated domains on a surface can, in principle, act as quantum bits or memory bits.[15,16] As a prerequisite to patterning, one must be able to fabricate uniform, thin films. Over the last few years, several approaches to making $Mn_{12}$-acetate films have been reported.[2,3,17-19] In addition, recent work in our laboratories has revealed that $Mn_{12}$-acetate films can be prepared from evaporation of solutions of the compound on $Si/SiO_2$ substrates by the dip and dry method.[20] AFM and XPS indicate that the films formed by this simple technique are of high quality and stable under ambient conditions.

In this paper, we report on efforts to pattern the $Mn_{12}$-acetate films prepared by the solution method. The approach we have devised integrates two features namely (1) forming $Mn_{12}$-acetate film by a solvent-evaporation technique and (2) the patterning of these films by photolithography. The micron-sized patterned films have been characterized by AFM and XPS. Magnetic measurements on the films confirm that the $Mn_{12}$-acetate molecules are intact, which in combination with the simplicity of the approach, makes this a particularly appealing method for organizing the magnetic molecules in artificial patterns.

In order to extend the success of $Mn_{12}$-acetate film production to the fabrication of patterns by photolithography, it was necessary to establish that the $Mn_{12}$-acetate films are stable in the presence of the chemicals used in the photolithographic steps. To verify the stability, $Mn_{12}$-acetate film samples were sonicated in the photoresist remover (MICROCHEM, REMOVER PG) for 1 hour. These conditions are sufficiently extreme such that, if the sample is still intact after these conditions, one can expect that a sample dipped into the remover and agitated for less than two minutes to lift the minimally cross-linked resist pattern off will remain unchanged. In order to make this assessment, the treated films were compared to the virgin $Mn_{12}$-acetate films by the use of XPS. No significant change was observed in the experiment. Enhancing our conclusion, AC-susceptibility of fully dried residual $Mn_{12}$-acetate powder after the photoresist remover treatment for 1 hour shows results consistent with intact $Mn_{12}$-acetate molecules. Furthermore, since the $Mn_{12}$-acetate films are deposited from solution, the photoresist should be stable in the solvent. We observed that the negative photo resist, SU-8-2002, is unaffected in 2-propanol, the solvent of choice for the $Mn_{12}$-acetate film deposition in this study, after sonicating for 1 hour, a treatment which again far exceeds the experimental conditions during the lithographic process.

The details for the patterning are described as follows: The $Si/SiO_2$ substrate used to fabricate the pattern was sequentially sonicated in acetone and isopropanol for two minutes and then dried by a flow of $N_2$ gas. Next, a layer of the negative photoresist, SU-8-2002, was spin-coated onto the substrate at 3000 rpm for 60 sec and then dried in air, resulting in a ~1200 nm thick resist layer. The photoresist layer was then exposed to uniform UV light through a photomask for 50sec, using a mask aligner. After development of the photoresist in SU-8 developer (MICROCHEM), deionized water was sprayed on the surface of the sample to rinse the sample. Next, the prepared solution ($2\times10^{-4}$ mol·L$^{-1}$ of Mn$_{12}$-acetate in 2-Propanol) was added dropwise to the patterned surface using a pipette. The substrate was tilted carefully from side to side to render the droplet flat on the surface and then the liquid film was dried in air. The deposition process was repeated five times to enhance the thickness of Mn$_{12}$-acetate on the film pattern. Finally, a lift-off of the resist material from the substrate, using photoresist remover (MICROCHEM), resulted in the Mn$_{12}$-acetate film pattern. All the above procedures were carried out in a clean room environment at ambient temperature.

An optical micrograph of the generated pattern is shown in Fig. 1. It shows rectangular structures with width greater than 5 μm. To investigate its surface characteristics, one of the patterned structures was scanned with an AFM (Digital Instruments Nanoscope IIIa). The scanned data were obtained in the tapping mode using a silicon tip in ambient conditions. The AFM image and the height profile of the bars in the optical image (Fig. 1) with nominal width of 5 μm each is shown in Fig. 2. The Mn$_{12}$-acetate samples were deposited within the three bars of the image (Fig. 2(a)). The roughness of the surface within the bars is clearly enhanced, compared to the outside regions, where the resist prevented Mn$_{12}$-acetate adsorption. The outside regions show the same surface roughness as pure Si/SiO$_2$ substrates, as seen in the height profile (Fig. 2(c)). For the height profile the region inside the square in the image was rescanned at a larger scan size (Fig. 2(b)). The height difference between the middle of each bar and the Si surface was measured to be about 5 nm by AFM. This gives an indication that the film patterns are composed of approximately two to three Mn$_{12}$-acetate molecular layers. The inside region of the bar was scanned separately on a 1 × 1 µm$^2$ size, and the results showed similar roughness as compared to the film images obtained from a film sample made by the evaporation method but not lithographically patterned. From the AFM analysis, it can be concluded that the evaporation method combined with photolithography technique does not result in any significant reduction in Mn$_{12}$-acetate film thickness.

It is of interest to ask at this point why higher walls are formed along the edge of the patterns. One plausible explanation invokes the difference in surface tension between the photoresist and the bare Si substrate. As it dries, the Mn$_{12}$-acetate solution film forms a curved surface along the boundary between two different surfaces, which leads to an inhomogeneous evaporation rate. Therefore, liquid flows from regions with low evaporation rate to regions with high evaporation rate, thereby transporting the Mn$_{12}$-acetate solute to curved regions where it accumulates. This phenomenon is also known as the "coffee stain" effect.[21] Another plausible explanation for the edge effect arises from a meniscus of the solution forming along the step-like boundary of the photoresist. As the solvent evaporates in the dip-and-dry process, the larger abundance of solute per area in the meniscus region leads to a thicker Mn$_{12}$-acetate film. At this point we cannot distinguish which of the effects dominates.

In order to verify the chemical stability of Mn$_{12}$-acetate during the patterning process, the lithographically formed Mn$_{12}$-acetate pattern was investigated by XPS (Kratos Axis System) in the image mode as well as the spectrum mode. XPS measurements were carried out in vacuum (~2×10$^{-8}$ Torr) with an Al X-ray gun (1486.6 eV). Fig. 3(a) shows the intensity versus the binding energy, for the Mn$_{12}$-acetate film pattern shown in Fig. 1 acquired in the spectrum mode. The electron pass energy in the analyzer was set at 80 eV. The Mn $2p_{3/2}$ and Mn $2p_{1/2}$ peaks were observed at 642.5 eV and 654.5 eV, respectively. The observation of both Mn peaks in the XPS spectrum indicates the presence of intact $Mn_{12}$-acetate molecules[22] in the film pattern. Fig. 3(b) and (c) show an XPS image and an intensity profile of a three bar structure. Because of the limited resolution in the XPS image mode (40 µm), larger patterns were selected. The image was captured in the same vacuum environment for 10000 sec by scanning at the binding energy for the Mn $2p_{3/2}$ peak (due to its higher intensity) and setting the electron pass energy at 160 eV. In order to reduce the background signal, the image data were also acquired at 630 eV binding energy, and then subtracted from the Mn $2p_{3/2}$ peak image data. The energy 630 eV was chosen roughly on the background level in the spectrum data (Fig. 3(a)). Because the photoemitted electrons are scattered randomly in space, the pattern boundaries in Fig. 3(b) are diffuse and the line profile in Fig. 3(c) is rounded. The region including a three bar structure was also scanned at the binding energy for the Si2p peak (100.8 eV) under identical conditions; As expected, the image and the intensity profile from the scan was the inversion of the pattern from Fig. 3(b) and (c). The XPS data indicate that the $Mn_{12}$-acetate is intact after the lithographic patterning procedure and that $Mn_{12}$-acetate constitutes the observed pattern.

Magnetic measurements were carried out using a Quantum Design SQUID magnetometer MPMS-XL. When the $Mn_{12}$-acetate patterns on Si substrates (as studied by AFM and XPS) were used for SQUID measurements, the output signal from the thin $Mn_{12}$-acetate film patterns was indistinguishable from the $Si/SiO_2$ substrate background signal. To overcome this problem, small pieces of polyethylene were used as substrates to deposit the films. The output signal from the pure polyethylene piece (without $Mn_{12}$-acetate films) was negligible as verified by SQUID measurements. The polyethylene pieces were dipped into a $Mn_{12}$-acetate solution in 2-Propanol and then dried in air. In order to ensure sufficient aggregation of $Mn_{12}$-acetate material, highly concentrated solution ($5\times10^{-4}$ mol·L$^{-1}$ of $Mn_{12}$-acetate in 2-Propanol) was used and the same process of dipping and drying was repeated 50 times. We think that the magnetic properties of the film formed on the polyethylene are similar to the patterned films, as no other chemical modification was applied to fabricate the film pattern and the uniformity of the film was comparable to the pattern shown in Fig.1. As indicated earlier, the integrity of $Mn_{12}$-acetate is preserved after the use of the photoresist remover. AC-susceptibility data were measured for the $Mn_{12}$-acetate films on the plastic straw piece. The real ($\chi'$) and imaginary ($\chi''$) components of the susceptibility were plotted as functions of temperature (2 K - 11 K) for three selected frequencies (1 Hz, 25 Hz and 50 Hz) as shown in Fig. 4. The temperatures and shifts for the peak positions are in agreement with the results acquired from $Mn_{12}$-acetate powder samples.[23] The noise and scatter in the data are attributed to the film nature on a substrate as compared to an oriented powder sample and to the relatively small amount of material in case of the film sample. The magnetic measurements demonstrate that the films formed by this technique are indeed composed of $Mn_{12}$-acetate.

In summary, we have presented a simple and reliable technique to generate lithographically structured film patterns composed of the $Mn_{12}$-acetate molecular magnets. It has been demonstrated that the chemical integrity and magnetic properties of the molecule are conserved while they go through various processing steps. Our method can be extended to the patterning of submicron feature size by using more advanced lithography techniques. Moreover, this strategy can be extended to different molecular materials if a suitable solvent for the dip and dry film formation can be found.

This research was supported by the Texas Advanced Research Program (010366-0038-2001) and by the Welch Foundation (A.1585). KRD gratefully acknowledge the National Science Foundation support from Nanoscale Science and Engineering (NIRT) Grant (DMR-0103455), the Telecommunication and Informatics Task Force (TITF 2001-3) at Texas A&M University, DOE-DE-FG03-02ER45999 and the National Science Foundation for an Equipment grant to purchase a SQUID magnetometer (NSF-9974899). Use of the TAMU/CIMS Materials Characterization Facility and useful discussions with Dr. W. Lackowski and Ms. Y. Vasilyeva are also acknowledged.

**Figure captions**

Fig. 1. Optical micrograph of $Mn_{12}$-acetate film pattern generated by photolithography.

Fig. 2. (a) AFM line scan image of three bars. (b) Enlarged image from the box in image (a). (c) Height profile of the image (b) to allow for measurement of $Mn_{12}$-acetate film thickness.

Fig. 3. (a) XPS spectrum of the $Mn_{12}$-acetate film pattern. (b) XPS image captured with the binding energy for Mn $2p_{3/2}$ peak. (c) Intensity profile along the white line in the image (b).

Fig. 4. Real part ($\chi'$) and imaginary part ($\chi''$) of AC-susceptibility data for a $Mn_{12}$-acetate film deposited on a polyethylene piece as a function of temperature at selected frequencies.

**Figure 1**. K. Kim et al

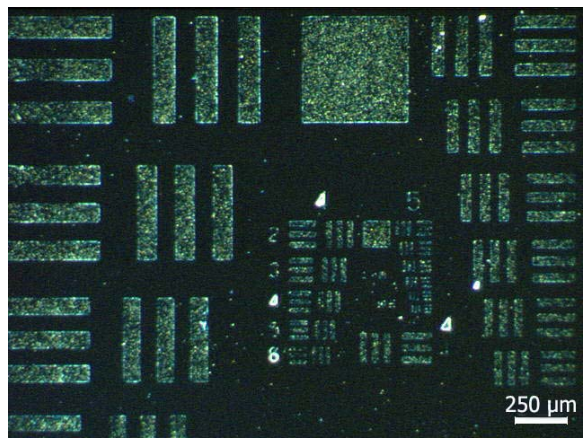

**Figure 2**. K.Kim et al.

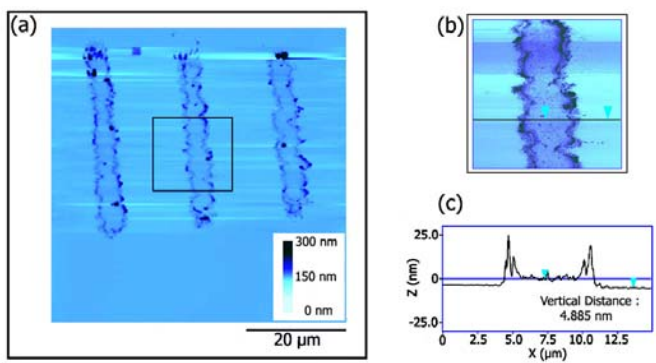

**Figure 3**. K.Kim et al.

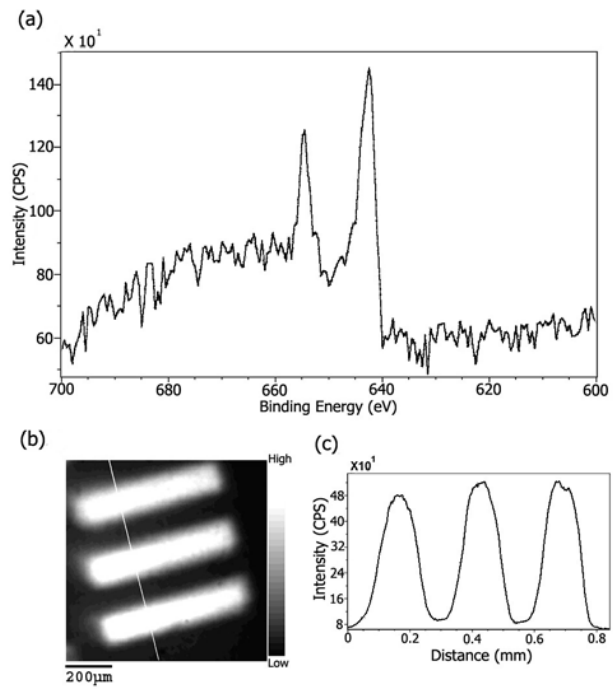

**Figure 4**. K.Kim et al.

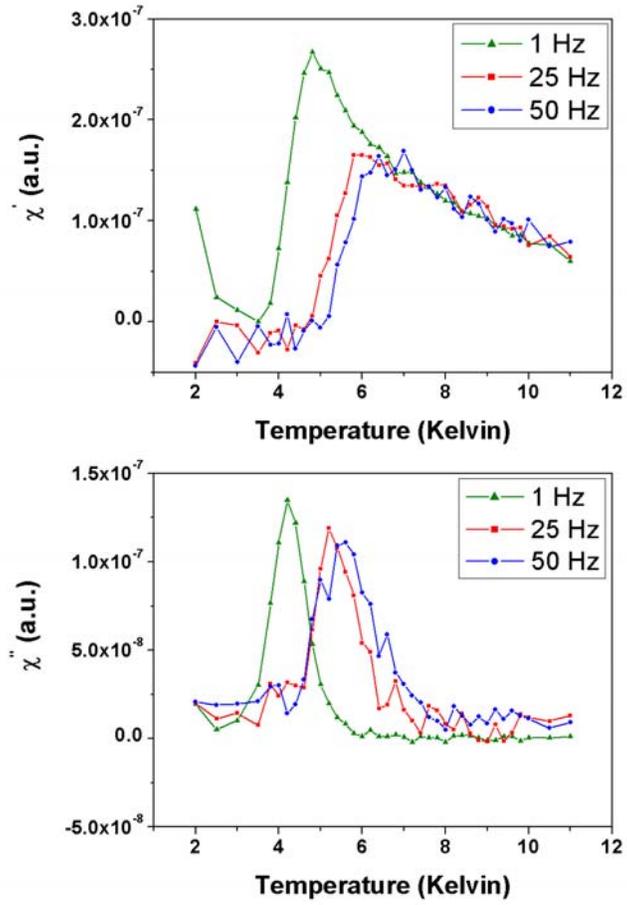